\documentstyle[prl,aps]{revtex}

\def\beginwide{
        \end{multicols} \vspace*{-0.5cm} \noindent
        \rule{3.5in}{.1mm}\rule{.1mm}{5mm} \widetext \medskip }
\def\endwide{
        \hspace*{3.5in}~\rule[-5mm]{.1mm}{5mm}\rule{3.5in}{.1mm}
        \begin{multicols}{2}\narrowtext \vspace*{-1.0cm} \noindent }
\def\beginwidetop{
        \end{multicols} \vspace*{-0.5cm} \noindent
        \widetext \medskip }
\def\endwidebottom{
        \begin{multicols}{2} \vspace*{-1.0cm} \noindent }

\begin{document}

\widetext

\title{Flat Spacetime Gravitation with a Preferred Foliation}
\author{J. Brian Pitts\thanks{Ilya Prigogine Center for Studies in Statistical
Mechanics and Complex Systems, Department of Physics, The University of Texas at Austin,
Texas USA 78712, jpitts@physics.utexas.edu} and W.C. Schieve\thanks{Ilya Prigogine Center for Studies in Statistical
Mechanics and Complex Systems, Department of Physics, The University of Texas at Austin,
Texas USA 78712} } 

\date{\today}

\maketitle


     Paralleling the formal derivation of general relativity as a flat spacetime theory, we
introduce in addition a preferred temporal foliation.  The physical interpretation of the formalism
is considered in the context of 5-dimensional
``parametrized'' and 4-dimensional preferred frame contexts.  In the former case, we suggest that our
earlier proposal of unconcatenated parametrized physics requires that 
the dependence on $\tau$ be rather slow.
In the 4-dimensional case, we consider and tentatively reject several areas of physics that
might require a preferred foliation, but find a need for one in the process
 (``flowing'') theory of time. 
We then suggest why such a foliation might reasonably be unobservable.


\section{Introduction}

     It is known that general relativity can be formally derived as a flat spacetime theory \cite{Gupta,Kraichnan,Thirring,Halpern,Feynman,OP,Deser,SliBimGRG}.  Such a derivation can be based on invariance (up to a boundary term) of the free gravitational action under
an infinitesimal gauge transformation and on universal coupling of all energy-momentum,
including that of gravitation itself, to serve as the source \cite{Deser,SliBimGRG}.  In this
context
there exists a gravitational energy-momentum tensor
\cite{Babak}, not merely a pseudotensor, so gravitational energy-momentum is localized 
in a coordinate-invariant (but gauge-variant) way.  As W. Thirring observed, it is not clear
\emph{a priori} why
Riemannian geometry is to be preferred over all the other sorts of geometry that exist, so a
derivation (as opposed to postulation) is attractive \cite{Thirring}.  
     
     Here we introduce on an $n$-dimensional space a preferred temporal foliation
$\partial_{\mu} \theta$ along side the $n$-dimensional flat metric $\eta_{\mu\nu}(x)$ (in
arbitrary coordinates) as nondynamical objects.  While the dimension $n$ is not specified, we
expect that $n = 4$ with metric signature $-+++$ and $n=5$ with signature $-+++-$
\cite{Fanchi,LandGreen,LandLorentz,LandParticles,Shnerb,LandShnerbGauge,HorwitzState,Saad,HorwitzPiron,Horwitz2Aspects,Lavie,Frastai,PittsPGR,PittsTau2} will be of the greatest
interest.  (If
$n=4$, then Greek indices run from 0 to 3 and $\theta = x^{0}= t$ for some preferred inertial
frame
in the natural coordinates.  If $n=5$, then Greek indices run over {0,1,2,3,5} with $\theta =
x^{5} = \tau$ in natural coordinates, $\tau$ being an invariant supertime.)  The foliation
$\partial_{\mu} \theta$ obeys $(\partial_{\mu} \theta) \eta^{\mu\nu} \partial_{\nu} \theta = -1$
and $\partial_{\nu} \partial_{\mu} \theta = 0.$  We define the unit normal covector by $n_{\mu}
= -\partial_{\mu} \theta$.  Its $\eta$-raised counterpart $n^{\nu} = n_{\mu} \eta^{\mu\nu}$ is
future-pointing \cite{MTW}.  The metric and foliation induce a projection tensor $h_{\mu\nu} =
\eta_{\mu\nu} + n_{\mu} n_{\nu}$.  If we agree to refer to $\theta$ as ``time'' (perhaps meaning
$\tau$) and the remaining dimensions as ``space'' (perhaps having signature $-+++$)
for convenience, then the projection tensor serves as the ``spatial'' metric.  The symmetric
gravitational potential is $\gamma^{\mu\nu}$ (of density weight 0), and bosonic matter fields are
denoted by $u$ (with all indices suppressed).  One could take $\gamma$ to
be either a contravariant density of any weight (except $\frac{1}{2}$) or a covariant density of
any weight (except $-\frac{1}{2}$).  (These weights are special because the resulting inverse
metric tensor density or metric tensor density, respectively, has a determinant of $-1$, so there are
only 9 independent components.  Thus, invertibility issues arise.)  In the massless case, this choice
makes no difference.  If it
were desired to add a gauge-symmetry-breaking mass term, the choice of
index position and weight would affect the results. 


\section{Spatial Gauge Invariance and Universal Momentum Coupling}

     This derivation is an improved version of previous work of ours on this subject
\cite{PittsTau2}, based on improvements in the corresponding derivation of general relativity
\cite{SliBimGRG}.  Unlike
our previous parametrized derivation \cite{PittsTau2}, this one does not assume that $n_{\mu}
n_{\nu} \gamma^{\mu\nu} = 0$, so one can retain the lapse function $N$ of an ADM split
\cite{MTW} as a nontrivial quantity, as opposed to requiring $N=1$ \emph{a priori} and thus not
varying it.  The assumption of gauge
invariance requires that the field be massless, except for the time-time part $n_{\mu} n_{\nu}
\gamma^{\mu\nu}$, which can be massive.  One can show this fact directly by writing the most
general algebraic quantity that is quadratic in the gravitational potentials.
(As in the case without a preferred foliation \cite{Freund}, there is no mathematical
objection to adding a mass term that breaks the gauge invariance.  A possible physical
difficulty with negative energy will be noted below.  Including both
mass terms would imply that the time-time part of the field had a different rest mass from the rest
of the field.)  


\subsection{Free Field Action}
     Let $S_{f}$ be the action for a free gravitational field (in the sense that the gravitational
coupling constant vanishes).
We require that $S_{f}$
change only by a boundary term under the infinitesimal gauge transformation 
\begin{eqnarray}
\gamma^{\mu\nu}
\rightarrow  \gamma^{\mu\nu}  +  
 \partial^{\mu} \xi^{\nu} + \partial^{\nu} \xi^{\mu}, 
\label{gaugeinv}
\end{eqnarray}
$\xi^{\nu}$ being a vector field obeying $\xi^{\nu} n_{\mu} = 0$.  

     For any $S_{f}$ invariant in this sense under (\ref{gaugeinv}), the free field equations'
divergence is purely temporal, for its spatial projection vanishes, as we now show.  The action
changes by 
\begin{eqnarray}
\delta S_{f} = \int d^{n}x  [\frac{\delta S_{f} }{ \delta \gamma^{\mu\nu} }  (\partial^{\nu}
\xi^{\mu} + \partial^{\mu} \xi^{\nu}) + e^{\mu},_{\mu}]= \int d^{n}x f^{\mu},_{\mu}.
\end{eqnarray}
The forms of the boundary terms $e^{\mu},_{\mu}$ and $f^{\mu},_{\mu}$
are not needed.  Integrating by parts,
letting $\xi^{\mu}$ have
compact support to annihilate the boundary terms (as we shall do throughout the paper), and
recalling the purely spatial character of $\xi^{\mu}$, we obtain the identity
\begin{eqnarray}
h^{\nu}_{\alpha} \partial^{\mu} \frac{\delta S_{f} } {  \delta \gamma^{\mu\nu} }  = 0.
\label{gaugeresult}
\end{eqnarray}


\subsection{The Asymmetric Metric Energy-Momentum Tensor}
    
     An expression for the total energy-momentum tensor can be derived from $S$ using the
metric recipe \cite{Kraichnan,Babak,Anderson} in the following way, making allowances for the
presence of another absolute object.  The action depends on the flat metric
$\eta_{\mu\nu}$, the foliation $n_{\mu}$, the gravitational potential $\gamma^{\mu\nu}$, and
bosonic matter fields $u$, the last representing any number of tensor densities of arbitrary weights
and index positions.
Under an arbitrary infinitesimal coordinate transformation described by a vector field
$\psi^{\mu}$, the action changes by the amount 
\begin{eqnarray}
\delta S = \int d^{n}x (\frac{\delta S}{\delta
\gamma^{\mu\nu}} \pounds_{\psi} \gamma^{\mu\nu}  +
\frac{\delta S}{\delta u} \pounds_{\psi} u +
\frac{\delta S}{\delta\eta^{\mu\nu} } \pounds_{\psi} \eta ^{\mu\nu} +
\frac{\delta S}{\delta n_{\mu} } \pounds_{\psi} n_{\mu}  + g^{\mu},_{\mu} ),
\end{eqnarray}  
where $\pounds_{\psi}$ is the Lie derivative with respect to $\psi^{\mu}$ and $g^{\mu},_{\mu} $ is another boundary term of no present interest.  But $S$ is a scalar,
so $\delta S= 0$.  Letting the matter and gravitational
field equations hold gives 
\begin{eqnarray}
\delta S = \int d^{n}x (\frac{\delta S}{\delta\eta^{\mu\nu} } \pounds_{\psi} \eta^{\mu\nu} +
\frac{\delta S}{\delta n_{\mu} } \pounds_{\psi} n_{\mu}  ) = 0,
\label{conserve}
\end{eqnarray}
or, upon rewriting the Lie derivatives in terms of the covariant derivative compatible with the flat
metric \cite{Wald}, 
\begin{eqnarray}
\partial^{\mu} \frac{\delta S}{\delta \eta^{\mu\nu} } - \frac{1}{2} n_{\nu} \partial_{\mu}
\frac{\delta S}{\delta n_{\mu } }  = 0.
\end{eqnarray}
One can identify $\frac{\delta S}{\delta \eta^{\mu\nu} } - \frac{1}{2} n_{\nu} \eta_{\mu\alpha}
\frac{\delta S}{\delta n_{\alpha}}$ as the energy-momentum tensor.  Its asymmetry reflects the
preferred character of $n_{\mu}$.  If we take the spatial projection of this quantity, we obtain the
tensor of momentum density and its flux $ h^{\nu}_{\alpha} \frac{\delta S}{\delta \eta^{\mu\nu}
}$. 


\subsection{Full Universally-Coupled Action}

     For the full theory, we postulate that the spatial projection of the Euler-Lagrange
equations should be just the spatial projection of the free field equations for $S_{f}$ augmented
by the momentum
tensor (including gravitational momentum):
\begin{eqnarray}
h^{\nu}_{\alpha}  \frac{\delta S}{\delta \gamma^{\mu\nu} } = h^{\nu}_{\alpha}   \frac{\delta
S_{f} }{\delta \gamma^{\mu\nu} } - \lambda  h^{\nu}_{\alpha} \frac{\delta S}{\delta
\eta^{\mu\nu} }, 
\label{Universal}
\end{eqnarray}  
where $\lambda$ is a coupling constant.  If $n=4$ and the theory in question is general relativity,
then $\lambda = - \sqrt{32 \pi G}$; for other theories, the Newtonian limit would need
consideration.  If $n=5$, then $\lambda$ has different dimensions from Newton's constant $G$,
with an additional length entering \cite{PittsPGR,CookI}.  Horwitz \emph{et al.} have previously
noted the entrance of an additional length in their parametrized electromagnetism
\cite{Horwitz2Aspects}.

     One is free to make a change of variables in
$S$ from the flat metric $\eta^{\mu\nu}$ and gravitational potential $\gamma^{\mu\nu}$ to
$\eta^{\mu\nu}$ and $g^{\mu\nu}$, where
\begin{eqnarray}
g^{\mu\nu} = \eta^{\mu\nu}  + \lambda \gamma^{\mu\nu}.
\end{eqnarray}  Equating coefficients of the variations gives
\begin{eqnarray}
 \frac{\delta S}{\delta \eta ^{\mu\nu}} |\gamma =   \frac{\delta S}{\delta
\eta^{\mu\nu}} |g  +  \frac{\delta S}{\delta g^{\mu\nu}}  
\end{eqnarray}
and 
\begin{eqnarray}   
 \frac{\delta S}{\delta \gamma
^{\mu\nu}}  =   \lambda \frac{\delta S}{\delta g ^{\mu\nu}}. 
\label{ELVarChange}
\end{eqnarray}
Putting these two results together gives
\begin{eqnarray}
 \lambda \frac{\delta S}{\delta \eta^{\mu\nu}} |\gamma =  \lambda   \frac{\delta S}{\delta
\eta^{\mu\nu}} |g + \frac{\delta S}{\delta \gamma^{\mu\nu}}.  
\label{StressSplit}
\end{eqnarray} 
Equation (\ref{StressSplit}) splits the energy-momentum tensor into one piece that vanishes when
the gravitational Euler-Lagrange equations hold and one piece that does not.  Using this result in
(\ref{Universal}) gives 
\begin{eqnarray}
\lambda h^{\nu}_{\alpha} \frac{\delta S}{\delta \eta^{\mu\nu}} |g = -h^{\nu}_{\alpha} 
\frac{\delta S_{f}}{\delta \gamma
^{\mu\nu}},
\label{Key}
\end{eqnarray}
which says that the spatial projection of the free field Euler-Lagrange derivative must equal (up to
a constant factor) that part of the momentum
tensor that does not vanish when the gravitational field equations hold.  Recalling
(\ref{gaugeresult}), one derives 
\begin{eqnarray}
h^{\nu}_{\alpha} \partial^{\mu} \frac{\delta S}{\delta \eta^{\mu\nu}}|g = 0, 
\label{projcurl}
\end{eqnarray}
which says that the part of the momentum tensor not proportional to the gravitational field
equations
has identically vanishing divergence. This result
concerning the splitting of the
momentum tensor will be used in considering the gauge transformations of the full theory.  It also
ensures that the gravitational field equations \emph{alone} entail conservation of momentum,
without any separate postulation of the matter equations.  (Previously the derivation
of a conserved momentum tensor required that gravity \emph{and matter} obey their field 
equations, as in (\ref{conserve}).)  But the conservation of energy still depends upon the matter
field
equations.

     Expanding the projection tensor gives
\begin{eqnarray}
\partial^{\mu} \frac{\delta S}{\delta \eta^{\mu\alpha}}|g + n^{\nu}n_{\alpha} \partial^{\mu}
\frac{\delta S}{\delta \eta^{\mu\nu}}|g  = 0.
\label{curl}
\end{eqnarray}
Let us divide the action into 3 pieces:  a piece $S_{1}$ independent of $\eta^{\mu\nu}$ (the
other variable being $g^{\mu\nu}$, not $\gamma^{\mu\nu}$), a piece $S_{2}$ that contributes a
symmetric curl (\emph{i.e.}, a
quantity with identically vanishing divergence on either index) to the stress tensor, and another
piece $S_{3}$.  We therefore write  
\begin{eqnarray}   
S = S_{1} [g, u, n] + S_{2}[g, u, \eta, n] + S_{3}[g, u, \eta, n],
\label{actionsplit}
\end{eqnarray}
suppressing all indices in the arguments.  Given the assumed property of $S_{2}$, one can write 
\begin{eqnarray}   
 -\eta^{\mu\alpha} \eta^{\nu\beta}  \frac{\delta S_{2} }{ \delta \eta ^{\alpha\beta}} |g =
\frac{1}{2}  \partial_{\rho} \partial_{\sigma} ( 
{\mathcal{M}}
^{[\mu\rho][\sigma\nu]} +   {\mathcal{M}}
^{[\nu\rho][\sigma\mu]} )  + b \sqrt{-\eta} \eta^{\mu\nu}
\end{eqnarray}
\cite{Wald} (pp. 89, 429), where ${\mathcal{M}} ^{\mu\rho\sigma\nu}$ is a tensor
density of weight $1$ and $b$ is a constant.  This result follows from the converse of
Poincar\'{e}'s lemma in flat spacetime.  
One easily verifies that if \cite{Kraichnan}  
\begin{eqnarray}   
S_{2} = \frac{1}{2} \int d^{n}x R_{\mu\nu\rho\sigma}(\eta)
{\mathcal{M}} ^{\mu\nu\rho\sigma} (g, u, \eta, n) + \int
d^{n}x {\mathcal{X}}^{\mu},_{\mu} + 2 b \int d^{n}x \sqrt{-\eta}, 
\end{eqnarray}
then $ \frac{\delta S_{2} }{ \delta \eta ^{\mu\nu}} |g $ has just the desired form, while $S_{2}$
does not affect the Euler-Lagrange equations because its value is $0$.  ${\mathcal{X}}^{\mu}$
is a weight $1$ vector density, because we
require that $S$ be a scalar.  For convenience we
deposit all boundary terms into $S_{2}$.

     One can also show that if
\begin{eqnarray}
S_{3} = -2 \int d^{n}x  \partial_{\beta} \psi^{\alpha\beta}(g, u, \eta, n) n_{\alpha}
\partial_{\mu} n^{\mu} , 
\end{eqnarray} then (\ref{curl}) is satisfied.  Now both terms of that equation are used, not just
one as for $S_{2}$.  Here $\psi^{\alpha\beta}(g, u, \eta, n) $ is any (2,0) weight 1 tensor density
constructed from all the fields, dynamical and nondynamical.  It turns out that 
\begin{eqnarray} 
\frac{ \delta S_{3}}{\delta \eta^{\mu\nu}} = - n^{\rho} n_{\alpha} \eta_{\mu\nu}
\partial_{\beta} \partial_{\rho} \psi^{\alpha\beta}  + 2 n_{\alpha} n_{(\nu} \partial_{\mu)}
\partial_{\beta}  \psi^{\alpha\beta}.
\end{eqnarray}       
Combining these 3 terms gives the total action
\begin{eqnarray}   
S = S_{1} [g^{\mu\nu}, u, n_{\mu}] + \frac{1}{2} \int d^{n}x R_{\mu\nu\rho\sigma} (\eta)
{\mathcal{M}} ^{\mu\nu\rho\sigma}(g, u, \eta, n)  + 2 b \int d^{n}x \sqrt{|\eta|} +  \nonumber  
\\   -2 \int d^{n}x  \partial_{\beta} \psi^{\alpha\beta}(g, u, \eta, n) n_{\alpha} \partial_{\mu}
n^{\mu}  + \int d^{n}x  {\mathcal{X}}^{\mu}(g,u,\eta,n),_{\mu}.
\end{eqnarray}
One sees that $S_{1}$ can contain mass and self-interaction terms
for the time-time component of the gravitational potential, using $\sqrt{-g}$ to the first power
and any suitable function of $g^{\mu\nu} n_{\mu} n_{\nu} + 1$. 
With this action, one could write down a number of theories, including (of course) general
relativity.  One would still need to verify that any given theory made sense theoretically (such as
by having adequate
positive energy properties) and empirically, for the universal coupling principle does not address
such questions fully.


\subsection{Gauge Invariance}

     It is instructive to determine what has become of the original free field gauge invariance. 
The scalar character of the action $S[g, u, \eta, n]$ implies that the variation of the action under a coordinate transformations vanishes:  \begin{eqnarray}
\delta S_{ct} = \int d^{n}x (\frac{\delta S}{\delta
g^{\mu\nu}} \pounds_{\psi} g^{\mu\nu}  +
\frac{\delta S}{\delta u} \pounds_{\psi} u +
\frac{\delta S}{\delta\eta^{\mu\nu} }|g \pounds_{\psi} \eta ^{\mu\nu} + h^{\mu},_{\mu} ) = 0.
\end{eqnarray} 
 But in a flat spacetime theory, invariance under coordinate
transformations is trivial.  A \emph{gauge} transformation, on the other hand, would be a
transformation that changes the action changes only by a boundary term, but is not a coordinate
transformation.  Using the coordinate transformation formula and noting that the terms involving
the absolute objects do not contribute more than a divergence, one easily verifies that a (pure)
gauge transformation is given
by $\delta
g^{\mu\nu} = \pounds_{\xi} g^{\mu\nu}$, $\delta u = \pounds_{\xi} u$, $\delta\eta^{\mu\nu}=
0$, $\delta n_{\mu}=
0$, where $\xi^{\mu} n_{\mu} = 0$, but $\xi^{\mu}$ is otherwise arbitrary.  In showing that the
term for the flat metric does not contribute nontrivially, one must recall from (\ref{projcurl})
above that
\begin{eqnarray}
h^{\nu}_{\alpha} \partial^{\mu} \frac{\delta S}{\delta \eta^{\mu\nu}}|g = 0
\end{eqnarray}
identically.  The term for the foliation does not contribute at all because $\pounds_{\xi} n_{\mu}
= 0$ on account of the constancy of $n_{\mu}$ and the assumption that $\xi^{\mu} n_{\mu} =
0$.  Thus, gauge transformations change (bosonic)
dynamical fields in the same way that `spatial' coordinate transformations do, but leave the
nondynamical
objects unchanged.  

\subsection{Bianchi Identities}

     Taking the independent variables to be those in $S[g^{\mu\nu}, u, \eta^{\mu\nu},
n_{\mu}]$, one can easily derive the Bianchi identities.  Letting the Euler-Lagrange equations
hold, one finds an additional equation that holds as a consequence.  This phenomenon
is not unprecedented: in unimodular general relativity (which also has a nondynamical object
present, namely, a volume element), this additional equation restores the trace of the Einstein
equations (up to an arbitrary cosmological constant) that failed to
appear in the Euler-Lagrange equations \cite{Unruh UGR}.  So a constraint not admitted through
the front door might still reenter through the back door.  

     From a Hamiltonian point of view, one
says that dynamical preservation of the momentum constraints (roughly, the time-space Einstein
equations) implies the Hamiltonian constraint (roughly, the time-time component of the Einstein
equations),
up to an arbitrary constant, in the unimodular theory.  On the other hand, we find that in the
theory obtained from the Hamiltonian for general relativity but with $N=1$ \emph{a priori} (so
the Hamiltonian constraint does not follow from the variation of the Hamiltonian), no further
constraint,
including any portion of the Hamiltonian constraint, is required to preserve the momentum
constraints. At least that is the case in the vacuum theory; we expect that minimally coupled
matter would behave similarly.  Also see (\cite{UnruhWald}).  But it would appear that this
theory
would have trouble with the negative energy degree of freedom constituted by the determinant of
the curved spatial metric, so it does not seem physically viable.
     
     We return now to the Lagrangian formulation for theories with a flat background metric and a
preferred foliation.  Making a coordinate transformation, letting the matter and gravitational field
equations
hold, discarding the boundary terms,
and using the arbitrariness of the coordinate transformation yields the relation \begin{eqnarray}
\partial^{\mu} \frac{\delta S}{\delta \eta^{\mu\nu} }|g - \frac{1}{2} n_{\nu} \partial_{\mu}
\frac{\delta S}{\delta n_{\mu } }  = 0.
\end{eqnarray}  On account of (\ref{projcurl}), the spatial projection of this equation already
holds identically.  The remainder, the temporal component, takes the form 
\begin{eqnarray}
2 n^{\nu} \partial^{\mu} \frac{\delta S}{\delta \eta^{\mu\nu}}|g  + \partial_{\mu} \frac{\delta
S}{\delta n_{\mu} }= 0.
\end{eqnarray}
When the action is expanded as using $S = S_{1} + S_{2} + S_{3}$, as in (\ref{actionsplit}),
several terms vanish, namely,
$\frac{ \delta S_{1} }{\delta \eta^{\mu\nu} }|g$, 
$\partial^{\mu} \frac{ \delta S_{2} }{\delta \eta^{\mu\nu} }|g$, and $\frac{ \delta S_{2} }{\delta
n_{\mu} }$.  After showing that $\frac{ \delta S_{3} }{\delta n_{\mu} } =
2 \partial^{\mu} n_{\alpha}  \partial_{\beta} \psi^{\alpha\beta}$, one finds that the contributions
from $S_{3}$ cancel each other.  The final result takes the rather simple form
\begin{eqnarray}
\frac{\partial}{\partial x^{\mu}} \frac{\delta S_{1} }{\delta n_{\mu } }  = 0.
\end{eqnarray}
The physical meaning of this equation will depend on the precise form of $S_{1}$.

 
\section{Interpretation of the $-+++-$ Formalism}

     In 5-dimensional form this work suggests itself as a route to a theory 
of ``parametrized'' gravitation, the extra dimension being the invariant supertime $\tau$.  If the
extra time dimension is to be interesting, it is necessary that physical fields be permitted to depend
on $\tau$ \cite{Horwitz2Aspects}.
But then one faces the question of relating a 5-dimensional description to the observed 4
dimensions.
One common approach in an electromagnetic context has been ``concatenation,'' in which the
$\tau$-dependent vector potential is integrated over all $\tau$
(from eternity past to eternity future) to give standard Maxwell potentials, the latter supposedly
being tied to
experiments \cite{Horwitz2Aspects,Lavie,LandGreen,LandLorentz,Shnerb,HorwitzState,Saad}.  
 We have previously argued that concatenation is unsatisfactory \cite{PittsPGR}, \emph{pace}
(\cite{Horwitz2Aspects}).  One reason is that it makes essential use of
the linearity of the field equations, but this linearity does not hold generically; in particular,
it is violated by any reasonable theory of gravity \cite{PittsPGR}.  It is doubtful that any variant
of concatenation of a nonlinear parametrized theory would give a plausible nonlinear
nonparametrized theory.  Even Yang-Mills theories, though formally similar
to electromagnetism, cannot be concatenated.
Another reason is that observations are said to be influenced by all values of $\tau$, including
future ones, yet $\tau$ is thought to be related to the
process/flowing aspect of time \cite{Horwitz2Aspects}.  The latter fact implies that experiments
performed by real people in ordinary life ought to occur at some definite moment (or finite
interval) of $\tau$.  So concatenation introduces the paradox of
backwards causation in $\tau$ \cite{PittsPGR}.  It also introduces a curious distinction between
measurement and evolution, like certain versions of quantum mechanics, rather than regarding
measurement as a specific kind of evolution.  

     Motivated by these criticisms, we previously suggested that omitting concatenation and
interpreting all experiences as involving a convective derivative with respect to $\tau$ along a
worldline might yield an adequate interpretation of the parametrized formalism \cite{PittsPGR}.
However, the non-concatenated view has a
drawback of its own, for such a theory
generically agrees with standard well-confirmed 4-dimensional electromagnetism (\emph{e.g.},
light speed measurements) only if the dependence on $\tau$ is quite weak \cite{Qualifier}.  This
limit is similar to the zero mode limit considered by Frastai and Horwitz \cite{Frastai}, who have
been aware of some of the difficulties with concatenation \cite{HorwitzEmail}.  They observed
that the zero mode limit is a sufficient condition for agreement of the parametrized theory with
experiment.  But our suggestion is that approaching the zero mode limit is a \emph{necessary}
condition.  For example, the lack of observed dispersion in light propagation indicates that the
$\tau$
derivatives
are much smaller than the $t$ derivatives in such contexts.  One could wonder
wonder what physical purpose $\tau$ serves \cite{Qualifier}.  If $\tau$ were to be
associated with temporal becoming only, with no physical content
(\emph{c.f.} (\cite{Miller}), then perhaps a more attractive and economical solution
exists in the $n=4$ context, as we will explain below. 

     A recent development in membrane theory might possibly be of interest here. Recently the 
somewhat
analogous question ``Can there be `large' extra dimensions?'' has received a surprising positive
answer.  The Randall-Sundrum scenario \cite{RandallSundrum,RSPhysToday} appears
to permit higher-dimensional theories to have large extra dimensions while giving empirically
reasonable results.  Perhaps one can imagine a parametrized analog of this move.

     Our discussion above has considered only parametrized field theory.
It is also worthwhile to consider the interpretation of parametrized particle mechanics, a subject
that has received some attention \cite{HorwitzPiron,PironReuse,Horwitz2Aspects,Trump}.
If the Stueckelberg-Feynman notion of positrons as electrons moving backwards-in-$t$ with
respect to $\tau$ is employed, as has often been suggested,
then the issue of backwards-in-$\tau$ causation appears.  One might instead require that particle
trajectories
be such that $t$ is always an invertible function of $\tau$ to avoid that problem.  One drawback is
that one loses the opportunity to consider pair-creation and pair-annihilation processes in a
classical framework. But then it appears that field theory is necessary after all to handle a variable
number of particles, and the problems above must be faced.

     In conclusion, it is clear that the parametrized formalism still faces fundamental
interpretive questions. 

     We also note some interesting work by Cawley \cite{Cawley} and by Hsu and Shi
\cite{HsuShi}.  The classical work of J. L. Cook \cite{CookI,CookGR}
will be considered in an appendix.


\section{Interpretation of the $-+++$ Formalism}

     In a 4-dimensional context this formalism corresponds to the existence of a preferred
reference frame.  It is generally assumed that no such thing exists, though the subject has received
some attention \cite{Selleri,Schmelzer}. The presumed nonexistence of a preferred frame is in
some important respects quite
helpful, because of the resulting tight restrictions on the number of theories that can
be conceived.  With a very few possible exceptions, all known physical processes are consistent
with the orthodox relativistic view that there is no preferred foliation (and that backwards-in-time
causation does not occur).  In view of the apparently limited gains and substantial losses realized
by giving up Lorentz invariance,
one might wonder what is the purpose of considering a preferred foliation in physics.  It would
seem that a rather good argument is needed to justify such work.  We now consider whether such
an argument is available.


\subsection{Possible Habitats for a Preferred Foliation in Physics?}

     One apparent difficulty for standard Lorentz-invariant physics is the remarkable behavior
seen in certain quantum mechanical experiments, such as by Aspect \emph{et al.}, in which 2
particles in a suitable superposed state seem to be able to `communicate' superluminally.
Much of the physics community seems to believe that locality is doomed, and has given up on it,
at least when its mind is on quantum mechanics.  However, this would
be a tremendous loss, and so it ought not to be accepted unnecessarily \cite{FerreroSantos}.  

     What can be said in defense of locality?  We do not claim to give a comprehensive review,
but only provide two suggestions.  Evidently the detector efficiency loophole is still open
\cite{SzaboFine,Genovese}.  Szabo and Fine's model even works for experiments testing the
GHZ scenario \cite{SzaboFine}, a more recent and perhaps more potent threat to local hidden
variables than Bell's theorem. 
Detector efficiencies are still too low to close this loophole \cite{Genovese}.  One also knows
that
the
experiments violating the Bell inequalities are compatible with the orthodox relativity if one is
prepared
to embrace ``superdeterminism'' \cite{Brans,Durt3Int,DurtDice,DaviesGhost}, which violates the
inequalities by introducing correlations between the hidden variables and the detector settings.  By
positing a common cause for these correlations, one can preserve orthodox
relativity, the Aspect experiments notwithstanding.  Because the GHZ theorem involves similar
locality assumptions to those involved in Bell's theorem \cite{CRB}, we suspect that it can be
subverted analogously.  However, this view's demanding philosophical underpinnings, such as its
denial of (libertarian) free
will\footnote{Free will faces a potent long-standing conceptual objection that an action that isn't
fully caused is to that extent merely random and thus un-free \cite{Van
Inwagen,Kane,EdwardsWill}, so the denial of free will might be inevitable on other grounds.  If
so, then the entrance fee for superdeterminism will decrease.}
and evident need for an all-determining Agent to correlate the initial conditions of the world,
might limit its appeal (see, \emph{e.g.}, Bell's attitude 
\cite{Bell} (pp. 100-103, 110, 154)).\footnote{On the other hand, the 3 major monotheistic
traditions all have (or had) strands that affirm theological determinism: Pharisaic Judaism
\cite{Metzger}, Reformed/Calvinist Christianity, and Islam.
 That there might be a natural affinity here is suggested by the language (\emph{e.g.},
(\cite{Durt3Int}) about events being ``already `written in a book'.''  The resemblance to Psalm
139:16 (NASB) cannot be accidental:
\begin{verse}  Thine eyes have seen my unformed substance; \\And in Thy book they were all
written, \\The days that were ordained for me, \\When as yet there was not one of
them.\end{verse} }  
The detector efficiency loophole has also seemed
unappealing to some, such as Bell \cite{Bell} (p. 109).  However, it is at least worthwhile to show
that these strategies exist, because they show that even
in this peculiar aspect of quantum mechanics, nothing is presently known with certainty that
requires a preferred frame.  

     Another trouble spot for the usual relativistic view of time is quantum gravity's ``problem
of time'' \cite{IshamTime}, which consists in the \emph{prima facie} disappearance of time from
quantum versions of general relativity.  However, we suspect 
that the problem lies not in the lack of a particular preferred frame (a feature shared with special
relativity; the success of standard field theory in other contexts suggests that this feature is not at
fault), but in the lack of a preferred class of
inertial frames peculiar to the form of gauge invariance of general relativity.  That is, plausibly the
`fault' lies in how general relativity differs from special relativity (general covariance, \emph{i.e.},
lack of nondynamical objects), so one needn't add structures unknown to special relativity to
address the issue.
One expects that the problem of time would disappear if one suitably introduced a nondynamical
background metric into the equations of motion.  Adding a small rest mass to the theory would be
an obvious way to
implement this procedure (and thereby obtain a nonvanishing Hamiltonian), if the traditional
negative-energy
objection to massive gravity \cite{DeserMass} (appendix on ``ghost'' theories) can be overcome. 
M.  Visser has recently suggested that it can \cite{Visser}.  One might also prefer that the curved
metric respect the
flat background's null cone structure, a nontrivial condition that, to our knowledge, has not been
successfully
imposed in an attractive way.\footnote{This issue was considered for (massless) general relativity
by Penrose \cite{Penrose}.  Massive theories differ from the massless case in several ways:  (1)
the gauge-dependence
of the relation between the two null cones disappears with the gauge freedom; (2) the supposed
difficulty with long-range divergence between the curved and flat null geodesics disappears, and
(3) the relation between the two null cones will tend to depend on the specific mass term
employed.}

     Recently another suggested habitat for the violation of Lorentz invariance has appeared. 
T. Jacobson and D. Mattingly have suggested that there is ``reason to doubt exact Lorentz
invariance:  it leads to divergences in quantum field theory associated with states of arbitrarily
high energy and momentum.
 This problem can be cured with a short distance cutoff which, however, breaks Lorentz
invariance'' \cite{Jacobson}.  They then introduce an ``aether'' consisting of a dynamical unit
timelike
vector (or covector) field.  Their aether, being dynamical and failing in general to define a
preferred foliation (because the covector typically is not a gradient), differs from what we
consider.  Their divergence argument might give a good reason
to consider a preferred foliation, such as we have considered here.  But it seems premature to put
too much reliance on this proposal.

\subsection{A Preferred Foliation from the Process Theory of Time}

     If these areas of physics do not provide sufficiently strong evidence for the existence of an
observable preferred foliation in physics, then one might ask if there are extra-physical reasons for
considering a preferred temporal foliation in physics.  It so happens that in the
20th century's central debate in the philosophy of time \cite{Smith}, one of the two views, if
established, would show that a preferred foliation exists at the most fundamental level.  (Below
we will find authors arguing that if a preferred
foliation exists, then presumably it manifests itself in physics.)  This is the debate about the
objectivity or otherwise of temporal becoming, that is, the `flow' of time
\cite{Dorato,IshamPolk}.  Some physicists and philosophers, based to a large degree on the
influence of relativity \cite{IshamPolk}, incline toward the ``block universe'' view that
regards all moments of time as ultimately equal in status; the notions of past, present, and future
are regarded as illusory or mind-dependent.  If the block view (also known as ``stasis,''
``B-theory,'' or ``tenseless'') is correct, then all facts can in principle be displayed on a (single)
spacetime diagram.\footnote{If one is reluctant to make
statements such as ``$2 + 2 = 4$ in 1980,'' one can admit a class of timeless truths also, but that
seems unnecessary \cite{Chisholm}.  Facts such as ``I am John Perry'' \cite{Perry}, if they are 
nontrivial,
might not fit on a spacetime diagram, but since their temporal properties are not the problem, we
can set them aside.} On the other hand, if a
spacetime diagram (perhaps augmented by timeless mathematical and logical truths) cannot
display all facts, because some facts have temporal properties inconsistent with such a
representation, then
the ``process'' (``A-theory,'' ``tensed'') view that affirms objective becoming will be established.

     The process view of time receives some unexpected assistance from stasis advocate D. H.
Mellor, who wrote \cite{MellorRT} (pp. 4-5):
\begin{quote}
     The tenseless camp often offers only weak inducements to join it:  the relative simplicity of
tenseless logic, for example, or its consonance with relativity's unification of space and time.  But
tenseless time needs a stronger sales pitch than that.
Tense is so striking an aspect of reality that only the most compelling argument justifies denying
it:  namely, that the tensed view of time is self-contradictory and so cannot be true.''
\end{quote}  
Mellor claims to find this needed contradiction in McTaggart's paradox, but this claim is not
generally accepted and indeed appears to be false \cite{SmithLanguage}.  If McTaggart's paradox
fails to demonstrate a contradiction,
but Mellor's judgement is otherwise to be trusted, then the process view of time wins already. 
But some will require a more compelling case, which we believe can be made.

     There is an argument (largely from (\cite{Craig}; a more developed version is forthcoming
\cite{CraigBooks12,CraigBooks34}) that appears to disprove the block
view by showing that it cannot accommodate certain facts \cite{PriorThank,PriorFormalities}. 
In order to make one's appointments on time, one frequently needs to know what time it is
\emph{now}.  For example, if one wants to pay taxes to the American government in a timely
way, one might want to know that ``It is now April 2.''  Otherwise, one might file many weeks or
even years late, because one just would not know when to file \cite{Perry}.  This sort of fact,
along with more general facts about what is occurring \emph{now}, cannot be represented
tenselessly \cite{Lewis,Wolterstorff}, such as on a spacetime diagram, or known by a timeless
being (even a divine one, as Kretzmann has in mind) \cite{Kretzmann}, \emph{i.e.}, one lacking
temporal location and
duration.
Let us see why this is the case. 
On such a diagram, one might make a mark at ``April 2'' on the time axis,
but this mark will soon be outdated, so it will no longer represent ``now.''  Trying to keep the
``now'' mark current would require continually erasing and drawing on the spacetime diagram,
which is of course illegal, for one then has a \emph{succession} of diagrams (a movie), not a
single one.  Neither will the fact ``It is April 2 on April 2'' be of any use, both because it
is a tautology \cite{Smith} and thus cannot inspire any action at all, and because it is always true
\cite{MellorTGTO} and thus cannot motivate action at any special moment.  If one finds the use
of a date label such as April 2 troubling (as if a substantivalist view of
time might be to blame), one could substitute some ordinary occurrence: ``A rooster is crowing
now'' would serve, provided
that the rooster only crows shortly before tax day.  So ``now'' points to one or more facts that the
block universe cannot accommodate.  We must therefore reject the following claim by H.
Reichenbach \cite{Reichenbach} (pp. 16,17):
\begin{quote}
     There is no other way to solve the problem of time than the way through physics \ldots If
time is objective the physicist must have discovered that fact, if [sic] there is Becoming the
physicist
must know it \ldots It is a hopeless enterprise to search for the nature of time without studying
physics.  If there is a solution to the philosophical problem of time, it is written down in
the equations of mathematical physics.
\end{quote} 
We must further differ from Reichenbach, who asserted that determinism would  exclude
becoming \cite{Reichenbach}, for irreducibly tensed facts provide a ground for an objective flow
of time, even if determinism is true.  (Interestingly, Reichenbach concluded that physics in fact
does ground time flow.)  Thus, we conclude that a preferred foliation generated by the ``moving
now'' exists.  
     

\subsection{Must a Preferred Foliation Be Physically Observable?}

     It seems natural to assume that if a preferred foliation exists, it ought to manifest itself in
physics fairly readily.\footnote{In the positivist era of the 20th century, the question in the the title
of this subsection might have received the answer ``of course,'' because the verificationist criterion
for meaning would have said that it was meaningless to talk about entities that are unobservable in
principle.  But such thought-stopping replies need not detain us today.} Such an intuition
indicates that Reichenbach's claim, though too strong, was not wholly
misguided.  In contemplating the notion of ``beables'' for quantum field theory, J. S. Bell wrote:
\begin{quote}
As with relativity before Einstein, there is then a preferred frame in the formulation of the theory
\dots but it is experimentally indistinguishable.  It seems an eccentric way to make a
world.\end{quote} \cite{Bell} (p. 180, ellipses in the
original; see also p. 155).  (This seems to have been Bell's \emph{a priori} judgement of the idea. 
While he thought it somewhat odd, he nevertheless thought it worthwhile to consider as ``the
cheapest resolution'' of what he saw \emph{a posteriori} to be a real difficulty posed by the
Aspect experiments \cite{DaviesGhost}.) Philosopher T. Maudlin, in attempting to make sense of
quantum mechanics and reconcile it to relativity, suggests that backwards causation or a preferred
reference frame might be the least unacceptable ways of doing so \cite{Maudlin}.  Concerning
the possibility of a preferred frame in making sense of quantum mechanics, Maudlin, perhaps
having in view Einstein's line that God is subtle but not malicious, writes:   
\begin{quote}
One way or another, God has played us a nasty trick.  The voice of Nature has always been faint,
but in this case it speaks in riddles and mumbles as well.  Quantum theory and
Relativity seem not to directly contradict each other, but neither can they easily be reconciled. 
Something has to give: either Relativity or some foundational element of our
world-picture must be modified \ldots the real challenge falls to the theologians of physics, who
must justify the ways of a deity who is, if not evil, at least extremely
mischievous.\cite{Maudlin} (p. 242) \end{quote}

     So if one is
persuaded that the flow of time is objective and that a preferred foliation ought to show itself
readily in physics, then one might consider the formalism above for gravitation with $n=4$, or
perhaps some other way of including the foliation in physics.
D. Bohm's nonlocal deterministic version of quantum mechanics is perhaps presently the most
vibrant work that assumes a preferred foliation of 4-dimensional spacetime; the
theory is presently being applied to quantum gravity and quantum cosmology
\cite{Shojai,GoldsteinBohm}.
(But its nonlocality is not easy to embrace, even if one can tolerate a preferred frame.)  However,
as Butterfield and Isham note, ``[m]ost general relativists feel [that] this response is too radical to
countenance:  they regard foliation independence as an undeniable
insight of relativity'' \cite{ButterfieldIsham,Rovelli}.  

     We suggest that the following explanatory strategy might relieve this tension between the
philosophical support for objective becoming and the dearth of physical support for a preferred
foliation, at least if theism is plausible.  Rather than regarding the inclusion of a physically invisible
(or nearly so) preferred foliation as an``eccentric'' \cite{Bell} or ``if not evil, at least extremely
mischievous'' \cite{Maudlin} way to make a world, one might suggest that the Maker rendered
the preferred foliation physically invisible as an act of \emph{benevolence} to physicists.  To put
it more plainly, the world was made Lorentz-invariant to make physics easier.\footnote{Another
suggestion might be that the world was made Lorentz-invariant to make physics prettier.}  As we
noted above, the requirement of Lorentz invariance so
restricts the possible theories that the principle of Lorentz invariance answers a vast number of
questions that would otherwise require laborious experimentation to settle.  Writing down,
\emph{e.g.}, all possible terms in linearized gravity, first given
Lorentz invariance, and then given a preferred foliation, would give one a clear sense of the
economy afforded by 
Lorentz invariance.  (Perhaps other symmetries are amenable to a similar interpretation.)  We will
content ourselves with the simpler scalar and vector field cases. 
Viewed from a spacetime perspective, rendering the
foliation invisible amounts to keeping the foliation from appearing nontrivially in the action, with
only
the flat metric present.  Viewed from the perspective of space-at-a-time, it means that the foliation
only appears in concert with the spatial metric,
in such a way that, along with a
fundamental constant with dimensions of velocity, neither the foliation nor the spatial metric
appears alone, but only the two combined into an effective spacetime metric.  

     We will now write down all possible terms that could appear in the Lagrangian density for
a real scalar field $\phi$, restricting the equations of motion to be linear and to have (at most)
second
derivatives.  The existence of
fundamental constants with dimensions of velocity ($c$) and angular momentum ($\hbar$) will be
assumed, although we choose units in which these constants have unit value.  Because the
equations of motion, which are perhaps more important
than the Lagrangian density, are unchanged by the addition of a divergence, we will regard terms
that are equal up to a divergence as equivalent to avoid overcounting, and will drop terms that are
themselves divergences.  For brevity, we use a
vertical bar to denote the flat metric's
covariant derivative.  We also the overdot notation to indicate the time derivative $n^{\mu}
\partial_{\mu}$.  The possible terms lacking $n_{\mu}$ are $\phi_{|\mu} \phi^{|\mu}$ and
$\phi^{2}$.  The first term is so basic that we will assume that it must always be present, even if
the foliation appears in the theory.  The values of all other coefficients are relative to this one,
which depends on the normalization only.  (Our results thus cannot be criticized as inflated, being
instead perhaps a bit pessimistic.)  If $n_{\mu}$ is also present, then $\dot{\phi}^{2}$ is also
available. 
Thus there are 2 unspecified constants in the case with a preferred foliation observable, but there
is only 1 with it absent, in the case of a scalar field. 
This is a savings already, though a modest one.  The vector case will be more compelling.   

     In the case of a vector field, the possible terms in the Lorentz-invariant case, taking into
account the restrictions above are these:  $A^{\mu}_{|\nu} A_{\mu}^{|\nu}$,
$A^{\mu}_{|\mu} A^{\nu}_{|\nu}$, $A^{\mu} A_{\mu}$, and $\epsilon^{\mu\nu\alpha\beta}
A_{\mu|\nu} A_{\alpha|\beta}$, leaving 3 unspecified
coefficients.  If Lorentz invariance is not required, there appear in addition
$\dot{A^{\mu}} \dot{A^{\nu}} n_{\mu} n_{\nu}$,    
$A^{\mu}_{|\sigma} A^{\nu|\sigma} n_{\mu} n_{\nu}$,
$\dot{A^{\mu}} \dot{A_{\mu}}$,
$\dot{A^{\mu}} A^{\nu}_{|\nu} n_{\mu}$,      
$A^{\mu} A^{\nu} n_{\mu} n_{\nu}$,
$A^{\mu}_{|\mu} A^{\nu} n_{\nu}$, and $\epsilon^{\mu\nu\alpha\beta} A_{\mu|\nu}
A_{\alpha} n_{\beta}$, giving 10 unspecified coefficients without Lorentz
invariance. 
The economy afforded by Lorentz invariance is thus considerable in the case of a vector field. 
One expects that it would be substantial for higher-rank fields as well.  We have not considered
complex fields or fermionic matter, but we imagine that
Lorentz invariance provides a respectable
simplification in those cases as well.  Complex fields have an interesting property not shared by a
single real scalar field, \emph{viz.}, they admit first-order-in-time equations of motion such
as the Schr\"{o}dinger equation.  (For a single real field, the most similar term in the Lagrangian
density,
$\phi \dot{\phi}$, is merely a divergence.)  So the investigation of these other types of fields, and
of sets of real fields with internal symmetry groups, might be of interest.
     
     In view of these simplifications, it is not too
implausible to think that temporal becoming is objective, and yet physics is exactly Lorentz
invariant, if
the existence of a benevolent God who supports the enterprise of physics is plausible.  Such a
divine motivation does fit naturally within traditional monotheistic religion.  According
to the
traditional story (Genesis 1:28), God tells the human race to reproduce 
and to fill the earth and subdue it, and to rule over other creatures.  This command, which has
been dubbed the ``cultural mandate,'' has been seen as encouraging the scientific study of natural
phenomena.
If this conclusion is correct,
then physicists and philosophers can pursue their respective visions of time without mutual
interference.  The great gulf that Stapp \cite{Stapp} and
Horwitz \emph{et al.} \cite{Horwitz2Aspects} have found between ``Einstein time'' and ``process
time'' is thus
perhaps not so fixed.  The need for a preferred foliation in physics would then
need to be shown on grounds other than the flow of time, if at all.  

     This suggestion just made also answers an objection against the idea that God is temporal,
\emph{i.e.}, has a location and a duration in time \cite{Wilcox,Ford,Gruenler,CraigRT}.  A
temporal God's knowledge of which events
are objectively simultaneous defines a preferred foliation.  But we have already addressed the
relativistic objection to a preferred foliation.  Introducing an omnipresent knower creates no
further difficulties.

     We close by concluding that even if one is persuaded of the existence of a preferred
foliation of spacetime, there presently seems to be no compelling reason for rejecting standard
Lorentz-invariant physics.  However, there are some reasons strong enough to make
consideration of the violation of Lorentz invariance an interesting pursuit.

\appendix
\section{$\tau$-related work of J. L. Cook}

     The work by J. L. Cook \cite{CookI,CookGR} deserves examination.  In view of its early
date for $\tau$-related work and its degree of development (to the point of including
numerical results), this work perhaps deserves a larger role in the history of parametrized
physics than it has hitherto received, if it is sound.  However, we will raise questions
about both the particle mechanics and the derivation of the 2-body gravitation interaction from a
field theory. 

     Concerning mechanics, there is a question about how is the constraint of unit 4-velocity is
preserved. Such a constraint is never mentioned by Cook, and it is suggested that particles have 4
degrees
of freedom \cite{CookI} (p. 122) \cite{CookGR} (p. 471). But given that Cook regards $\tau$ 
as proper time \cite{CookI} (p. 121), this constraint ought to hold. It is an important feature of
relativistic mechanics, so its omission is puzzling.

     In addressing the 2-body problem, Cook undertakes to derive the
2-body electromagnetic and gravitational interactions from corresponding
field theories.  For electromagnetism,
$\tau$-dependent potentials are
introduced in some postulated field equations.  The justification for the
equations seems to be not so much a principled $\tau$-dependent
theory (which might follow from an action principle),
but a desire to obtain equations that, upon integration over all $\tau$ (a
procedure later called ``concatenation'' by Horwitz \emph{et al.}, and which
Cook sees as implicit in the work of Wheeler and Feynman), yield
Maxwell's
equations.  While an extra scalar potential in some way corresponding to
$\tau$ is introduced, it seems to play little role in the theory.

     In view of the similarity between
electromagnetism and general relativity, one might expect that the search for the
gravitational 2-body interaction would follow the same plan as in electromagnetism.  Surprisingly
enough, that is not the case.  Study of Cook's work leaves us with
significant unanswered questions.  

     Regarding the field theories used to derive the 2-body potentials, why
is it that for gravitation, 5-dimensional
$\tau$-dependent Einstein field equations are assumed to hold\footnote{We now turn to amend a
previous criticism \cite{PittsPGR} of Cook's paper \cite{CookI} to
the effect that the allowed dependence on $\tau$ was restricted.  This remark was based on the
lack of $\tau$-derivatives in equation 46 of (\cite{CookI}),
based on the definition of the $\Box$ symbol in equation 42b.  However, from equation 44 and
the coordinate condition just above equation 46, it follows that the $\Box$ symbol in equation 46
ought to refer to a 5-dimensional wave equation, not a 4-dimensional one.  One could then regard
the lack of $\frac{\partial^{2}}{\partial \tau^{2}}$ in equation 46 as merely a
typographical error.}, but 5-dimensional Maxwell
equations are not assumed for electromagnetism \cite{CookI}?  This inconsistency  suggests a
degree of arbitrariness in the assumed field theories.  

          It should also be noted that the 5-dimensional Einstein equations are written in terms of
2-body \emph{relative}
coordinates, the center of mass motion presumably having been factored out.  But
what does it mean to write a field theory in terms of relative coordinates?  The relative
coordinates
in a 2-body problem take only 1 value at any moment (meaning a moment of $\tau$, in this case),
but the coordinates in a field theory
take on all values.  
Furthermore, on what grounds does one assume that something like 5-dimensional Einstein field
equations should hold with respect to the relative coordinates?  This is hardly follows from
``standard methods'' (p. 127) or M{\o}ller's book \cite{MollerBook}, to which Cook appeals. 
The 
situation becomes still less clear for the $n$-body problem, for then there are $n-1$ sets of
coordinates available.  
  
     Finally, concatenation of the linearized
gravitational potential plays an essential role in finding the line
element \cite{CookI}.  As we have noted above, concatenation of a $\tau$-local theory to
give an ordinary theory depends
essentially on the linearity of the
theory \cite{PittsPGR}, but linearity holds only in a weak-field approximation for
gravitation. What is the generalization to the case of the full nonlinear Einstein equations that are
assumed?  Cook gives no hint of an answer, and we cannot think of any attractive candidates
either.

     In view of the questions listed above, we conclude that, although ``interactions over surfaces
of equal proper times are not
known and must be determined from field equations'' \cite{CookI}, the gravitational interaction
found in (\cite{CookI,CookGR}) has not been persuasively derived from a field theory.  
     

\acknowledgments

     We thank Dr. L. P. Horwitz for many helpful comments on this paper.



\begin{thebibliography}{10}

\bibitem{Gupta} S. Gupta, ``Gravitation and electromagnetism,'' \emph{Phys. Rev.} {\bf 96}, 1683 (1954).

\bibitem{Kraichnan} R. Kraichnan, ``Special-relativistic derivation of generally covariant
gravitation theory,''  \emph{Phys. Rev.} {\bf 98}, 1118 (1955); also ``Possibility of unequal
gravitational and inertial masses,'' \emph{Phys. Rev.} {\bf 101}, 482 (1956).

\bibitem{Thirring} W. Thirring, ``An alternative approach to the theory of gravitation,''
\emph{Ann. Phys. (N.Y.)} {\bf  16}, 96 (1961).

\bibitem{Halpern} L. Halpern, ``On the structure of the gravitation self interaction,'' \emph{Bull.
Cl. Sci. Acad. R. Belg., 5e serie,} {\bf 49}, 226 (1963).

\bibitem{Feynman} R. P. Feynman, F. B. Morinigo, and W. G. Wagner, \emph{Feynman
Lectures
on Gravitation}, B. Hatfield, \emph{ed.} (Addison-Wesley, Reading, Mass., 1995).

\bibitem{OP} V. Ogievetsky and I. Polubarinov, ``Interacting field of spin 2 and the Einstein
equations,'' \emph{Ann. Phys. (N.Y.)} {\bf 35}, 167 (1965).

\bibitem{Deser} S. Deser, ``Self-interaction and gauge invariance,'' \emph{Gen. Rel. Gravit.} {\bf
1} 9, (1970).

\bibitem{SliBimGRG}  J. B. Pitts and W. C. Schieve, ``Slightly bimetric gravitation,''
\emph{submitted}.

\bibitem{Babak} S. V. Babak and L. P. Grishchuk, ``The energy-momentum tensor for the
gravitational field,''  \emph{Phys. Rev. D} {\bf 61}, 024038 (2000).

\bibitem{Fanchi} J. R. Fanchi, \emph{Parametrized Relativistic Quantum Theory}. (Kluwer
Academic,
Dordrecht, 1993). 

\bibitem{LandGreen} M. C. Land and L. P. Horwitz, ``Green's functions for off-shell
electromagnetism and spacelike correlations,'' \emph{Found. Phys.} {\bf 21}, 299 (1991).

\bibitem{LandLorentz} M. C. Land and L. P. Horwitz, ``The Lorentz force and
energy-momentum
for off-shell electromagnetism,'' \emph{Found. Phys. Let.} {\bf 4}, 61 (1991).

\bibitem{LandParticles} M. C. Land, ``Particles and events in classical off-shell electrodynamics,''
\emph{Found.Phys.} {\bf 27}, 19 (1997).

\bibitem{Shnerb} N. Shnerb and L. P. Horwitz, ``Canonical quantization of four- and
five-dimensional U(1) gauge theories,''  \emph{Phys. Rev. A } {\bf 48}, 4068 (1993). 
 
\bibitem{LandShnerbGauge}  M. C. Land, N. Shnerb, and L. P. Horwitz, ``On Feynman's
approach to the foundations of gauge theory,'' \emph{J. Math. Phys.} {\bf 36}, 3263 (1995).

\bibitem{HorwitzState}  L. P. Horwitz, ``On the definition and evolution of states in relativistic
classical and quantum
mechanics,'' \emph{Found. Phys.} {\bf 22}, 421 (1992).

\bibitem{Saad}  D. Saad, L. P. Horwitz, and R. I. Arshansky, ``Off-shell electromagnetism in
manifestly covariant relativistic quantum mechanics,'' \emph{Found. Phys.} {\bf 19}, 1125
(1989).

\bibitem{HorwitzPiron}  L. P. Horwitz and C. Piron, ``Relativistic dynamics,'' \emph{Helv. Phys.
Acta } {\bf 46}, 316 (1973).

\bibitem{Horwitz2Aspects}  L. P. Horwitz, R. I. Arshansky, and A. C. Elitzur, ``On the two
aspects of time: the distinction and its implications,'' \emph{Found. Phys.} {\bf 18}, 1159 (1988).

\bibitem{Lavie}  R. Arshansky, L. P. Horwitz, and Y. Lavie, ``Particles \emph{vs.} events: the
concatenated structure of world lines in relativistic quantum mechanics,'' \emph{Found. Phys.}
{\bf 13}, 1167
(1983).

\bibitem{Frastai} J. Frastai and L. P. Horwitz, ``Off-shell fields and Pauli-Villars regularization,''
\emph{Found. Phys.} {\bf 25}, 1495 (1995).

\bibitem{PittsPGR} J. B. Pitts and W. C. Schieve, ``On parametrized general relativity,''
\emph{Found. Phys.} {\bf 28}, 1417 (1998).

\bibitem{PittsTau2} J. B. Pitts and W. C. Schieve, ``On the form of parametrized gravitation in
flat spacetime,''  \emph{Found. Phys.} {\bf 29}, 1977 (1999). 

\bibitem{MTW} C. Misner, K. Thorne, and J. Wheeler, \emph{Gravitation}, (Freeman, New
York, 1973).  

\bibitem{Freund} P. G. O. Freund, A. Maheshwari, and E. Schonberg, ``Finite-range gravitation,''
\emph{Ap. J.} {\bf 157}, 857 (1969).

\bibitem{Anderson} J. L. Anderson, \emph{Principles of Relativity Physics}, (Academic, New
York, 1967).  

\bibitem{CookI} J. L. Cook, ``Solutions of the relativistic two-body problem,''  \emph{Aust. J.
Phys.} {\bf 25}, 117 (1972).

\bibitem{Wald} R. Wald, \emph{General Relativity}, (Univ. Chicago, Chicago 1984).

\bibitem{Unruh UGR} W. G. Unruh, ``Unimodular theory of canonical quantum gravity,''
\emph{Phys. Rev. D} {\bf 40}, 1048 (1989).

\bibitem{UnruhWald} W. G. Unruh and R. M. Wald,``Time and the interpretation of canonical
quantum gravity,'' \emph{Phys. Rev. D} {\bf 40}, 2598 (1989).  

\bibitem{HorwitzEmail}  One of us (J. B. P.) thanks Dr. L. P. Horwitz for correspondence on this
matter.

\bibitem{Qualifier}  One of us (J. B. P.) thanks Drs. R. Matzner and M. Choptuik for making this
point and Dr. D. Salisbury for related thoughts.

\bibitem{Miller} P. Miller,``On `becoming' as a fifth dimension,'' in \emph{Physics and the
Ultimate Significance of Time},
D. R. Griffin, \emph{ed.} (SUNY, Albany 1986).

\bibitem{RandallSundrum}  L. Randall and R. Sundrum, ``An alternative to compactification,''
\emph{Phys. Rev. Lett.} {\bf 83}, 4690 (1999) .
 
\bibitem{RSPhysToday} ``Physics update: an alternative to compactification,''  \emph{Physics
Today}, no. 12, 9 (1999) .

\bibitem{PironReuse}  C. Piron and F. Reuse, ``The relativistic two body problem,'' \emph{Helv.
Phys. Acta} {\bf 48}, 631 (1975).

\bibitem{Trump}  M. A. Trump and W. C. Schieve, \emph{Classical Relativistic Many-Body
Dynamics}, (Kluwer Academic, Dordrecht, 1999).

\bibitem{Cawley}  R. G. Cawley, ``Observer theories based on Stueckelberg equations of
motion,''  \emph{Int. J. Theor. Phys.} {\bf 3}, 483 (1970).

\bibitem{HsuShi} J. P. Hsu and T. Y. Shi, ``Hamiltonians within the relativistic dynamics with a
scalar evolution variable,'' \emph{Phys. Rev. D} {bf 26}, 2745 (1982).

\bibitem{CookGR}  J. L. Cook, ``General relativity in the equal proper time formalism,'' 
\emph{Aust. J. Phys.} {\bf 25}, 469 (1972). 

\bibitem{Selleri} F. Selleri, ``The relativity principle and the nature of time,'' \emph{Found.
Phys.} {\bf 27}, 1527 (1997).

\bibitem{Schmelzer} I. Schmelzer, ``General ether theory,'' \emph{Los Alamos Preprints},
xxx.lanl.gov, gr-qc/0001101. 

\bibitem{FerreroSantos}  M. Ferrero and E. Santos, ``Empirical consequences of the scientific
construction: the program of local hidden-variables theories in quantum mechanics,''
\emph{Found. Phys.} {\bf 27}, 765 (1997). 

\bibitem{SzaboFine}  L. E. Szabo and A. Fine, ``A local hidden variable theory for the GHZ
experiment,'' \emph{Los Alamos Preprints}, xxx.lanl.gov,
quant-ph/0007102. 

\bibitem{Genovese}  M. Genovese, G. Brida, C. Novero, and E. Predazzi, ``Experimental test of
local realism using non-maximally entangled states,'' \emph{Los Alamos
Preprints}, xxx.lanl.gov, quant-ph/0009067.


\bibitem{Brans} C. H. Brans, ``Bell's theorem does not eliminate fully causal hidden variables,'' 
\emph{Int. J. Theor. Phys.} {\bf 27}, 219 (1988).

\bibitem{Durt3Int} T. Durt, ``Three interpretations of the violations of Bell's inequalities,''
\emph{Found. Phys.} {\bf 27}, 415 (1997).
 
\bibitem{DurtDice} T. Durt, ``Why God might play dice,'' \emph{Int. J. Theor. Phys.} {\bf 35},
2271 (1996).

\bibitem{DaviesGhost} J. Bell, in \emph{The Ghost in the Atom}, P. C. W. Davies and 
J. R. Brown, \emph{ed.} (Cambridge Univ., Cambridge 1986).

\bibitem{CRB}  R. K. Clifton, M. L. G. Redhead, and J. N. Butterfield, ``Generalization of the
Greenberger-Horne-Zeilinger algebraic proof of nonlocality,'' \emph{Found. Phys.} {\bf
21}, 149 (1991).

\bibitem{Bell}  J. S. Bell, \emph{Speakable and Unspeakable in Quantum Mechanics},
(Cambridge
Univ., Cambridge, 1987).  

\bibitem{Van Inwagen} P. Van Inwagen, \emph{An Essay on Free Will}, (Oxford Univ.,
Oxford, 1986).  Though an advocate of free will, Van Inwagen finds no way to answer this
``Mind'' argument.

\bibitem{Kane} R. Kane,  \emph{The Significance of Free Will}, (Oxford Univ., Oxford, 1998). 
Kane, a defender of free will, wrestles with this argument through much of the book.  

\bibitem{EdwardsWill} J. Edwards, \emph{The Freedom of the Will}, (Soli Deo Gloria,
Morgan, Pennsylvania, 1996).  Reprint of (Thomas Nelson, London, 1845).  

\bibitem{Metzger} B. Metzger, \emph{The New Testament:  Its Background, Growth, and
Content}, 2nd ed. (Abingdon, Nashville, 1983).  Metzger cites Josephus.

\bibitem{IshamTime}  C. J. Isham, ``Canonical quantum gravity and the problem of time,'' 
\emph{Los Alamos Preprints}, xxx.lanl.gov, gr-qc/9210011.

\bibitem{DeserMass} D. Boulware and  S. Deser, ``Can gravitation have a finite range?''
\emph{Phys. Rev. D} {\bf 6}, 3368 (1972).

\bibitem{Visser}  M. Visser, ``Mass for the graviton,'' \emph{Gen. Rel. Gravit.} {\bf 30}, 1717
(1998).

\bibitem{Penrose}  R. Penrose, ``On Schwarzschild causality---a problem for ``Lorentz covariant''
general relativity,'' in \emph{Essays in General Relativity---A Festschrift for
Abraham Taub}, \emph{ed.} F. J. Tipler, (Academic, New York, 1980).

\bibitem{Jacobson}  T. Jacobson and D. Mattingly, ``Gravity with a dynamical preferred frame,''
\emph{Los Alamos Preprints}, xxx.lanl.gov,
gr-qc/0007031.

\bibitem{Smith}  Q. Smith, ``Problems with the new tenseless theory of time,''
\emph{Philosophical Studies} {\bf 52}, 371 (1987). 

\bibitem{Dorato}  M. Dorato, \emph{Time and Reality:  Spacetime Physics and the Objectivity of
Temporal Becoming}, (CLUEB, Bologna, 1995).

\bibitem{IshamPolk}  C. J. Isham and J. C. Polkinghorne, ``The debate over the block universe,''
in \emph{Quantum Cosmology and the Laws of Nature:  Scientific Perspectives on Divine
Action},
2nd ed., \emph{ed.} R. J. Russell, N. Murphey, and C. J. Isham, \emph{ed.} (Vatican
Observatory,
Vatican City State, and the Center for Theology and
the Natural Sciences, Berkeley, 1999).

\bibitem{Chisholm}  R. M. Chisholm and D. W. Zimmerman, ``Theology and tense,''
\emph{No\^{u}s} {\bf 31}, 262  (1997). 

\bibitem{Perry}  J. Perry, ``The problem of the essential indexical,'' \emph{No\^{u}s} {\bf 13}, 3
(1979).

\bibitem{MellorRT}  D. H. Mellor,  \emph{Real Time}, (Cambridge Univ., Cambridge, 1981).

\bibitem{SmithLanguage}  Q. Smith, \emph{Language and Time}, (Oxford Univ., New York,
1993).

\bibitem{Craig}  W. L. Craig, lectures given at Talbott Seminary, Biola Univ. (1995).  The
influence of these lectures on one of us (J. B. P.) goes beyond the citations.

\bibitem{CraigBooks12}  W. L. Craig, \emph{The Tensed Theory of Time}, and \emph{The
Tenseless Theory of Time}, (Synthese Library, Kluwer Academic, Dordrecht, forthcoming). 

\bibitem{CraigBooks34}  W. L. Craig, \emph{Time and the Metaphysics of Relativity}, and
\emph{God, Time and Eternity}, (Kluwer Academic, Dordrecht, forthcoming).  

\bibitem{PriorThank}  A. N. Prior, ``Thank goodness that's over,'' \emph{Philosophy} {\bf 34},
12
(1959).

\bibitem{PriorFormalities}  A. N. Prior, ``The formalities of omniscience,'' \emph{Philosophy}
{\bf
37}, 114 (1962).

\bibitem{Lewis}  D. Lewis, ``Attitudes \emph{de dicto} and \emph{de se},'' \emph{The
Philosophical Review} {\bf 88}, 513 (1979).

\bibitem{Wolterstorff}  N. Wolterstorff, ``God everlasting,'' in \emph{Contemporary Philosophy
of Religion}, S. M. Cahn and D. Shatz, emph{ed.} (Oxford Univ., New York, 1982).

\bibitem{Kretzmann}  N. Kretzmann, ``Omniscience and immutability,'' \emph{The Journal of
Philosophy}  {\bf 63}, 409 (1966).

\bibitem{MellorTGTO}  D. H. Mellor, ```Thank goodness that's over', '' \emph{Ratio} {\bf 23},
20 (1981).  Mellor is attempting to refute this line of argument.

\bibitem{Reichenbach}  H. Reichenbach, \emph{The Direction of Time}, (Univ. California,
Berkeley, 1956).

\bibitem{Griffin} D. R. Griffin, \emph{Physics and the Ultimate Significance of Time}, (SUNY,
Albany, 1986).

\bibitem{Maudlin} T. Maudlin, \emph{Quantum Non-Locality and Relativity: Metaphysical
Intimations of Modern Physics}, (Blackwell, Oxford, 1994).

\bibitem{Shojai}  F. Shojai and M. Golshani, ``On the general covariance in Bohmian quantum
gravity,'' \emph{Int. J. Mod. Phys. A} {\bf 13}, 2135 (1998).

\bibitem{GoldsteinBohm}  S. Goldstein and S. Teufel, ``Quantum spacetime
without observers: ontological clarity and the conceptual foundations of quantum gravity,''
\emph{Los Alamos Preprints}, xxx.lanl.gov,
quant-ph/9902018, to appear in \emph{Physics Meets Philosophy at the Planck Scale},
C. Callender and N. Huggett, \emph{ed.} (Cambridge Univ., forthcoming).


\bibitem{ButterfieldIsham}  J. Butterfield and C. J. Isham, ``Spacetime and the philosophical
challenge of quantum gravity,''  \emph{Los Alamos Preprints}, xxx.lanl.gov,
gr-qc/9903072, to appear in \emph{Physics Meets Philosophy at the Planck Scale}, C. Callender
and N. Huggett, \emph{ed.}  (Cambridge Univ., forthcoming). 

\bibitem{Rovelli}  One of us (J. B. P.) thanks Dr. C. Rovelli for stimulating correspondence on
this issue.  

\bibitem{Stapp} H. P. Stapp, ``Einstein time and process time,'' in \emph{Physics and the
Ultimate
Significance of Time}, D. R. Griffin, \emph{ed.} (SUNY, Albany, 1986).

\bibitem{Wilcox}  J. T. Wilcox, ``A question from physics for certain theists,''  \emph{The Journal of Religion} {\bf 41}, 293 (1961).

\bibitem{Ford}  L. S. Ford, ``Is process theism consistent with relativity theory?''  \emph{The Journal of Religion} {\bf 48}, 124 (1968).

\bibitem{Gruenler}  R. G. Gruenler, \emph{The Inexhaustible God}, (Baker, Grand Rapids, 1983).  

\bibitem{CraigRT}  W. L. Craig, ``God and real time,'' \emph{Religious Studies} {\bf 26}, 335 (1990).

\bibitem{MollerBook}  C. M{\o}ller, \emph{The Theory of Relativity}, (Clarendon, Oxford, 1952).


\end{thebibliography}
\end{document}